\documentclass{article}

\usepackage[square,numbers]{natbib}

\usepackage[final]{neurips_2023_ml4ps}

\usepackage[utf8]{inputenc} 
\usepackage[T1]{fontenc}    
\usepackage{hyperref}       
\usepackage{url}            
\usepackage{booktabs}       
\usepackage{amsfonts}       
\usepackage{nicefrac}       
\usepackage{microtype}      
\usepackage{xcolor}         
\usepackage{subfiles}
\usepackage{amsmath} 
\usepackage{graphicx}
\usepackage{bm}
\usepackage{caption}
\usepackage{subcaption}
\usepackage{float}

\title{Generating Multiphase Fluid Configurations in Fractures using Diffusion Models}

\author{%
Jaehong Chung$^{1, 3*}$ \quad Agnese Marcato$^{2, 3}$ \quad Eric J. Guiltinan$^3$ \\
\textbf{Tapan Mukerji}$^1$ \quad \textbf{Yen Ting Lin}$^3$ \quad \textbf{Javier E. Santos}$^3$\\\\
$^1$Stanford University \quad $^2$Politecnico di Torino \quad $^3$Los Alamos National Laboratory\\
\texttt{\{jhchung1, mukerji\}@stanford.edu}\\
\texttt{\{amarcato, eric.guiltinan, yentingl, jesantos\}@lanl.gov}\\
}

\begin{document}
\maketitle

\begin{abstract}
Pore-scale simulations accurately describe transport properties of fluids in the subsurface. These simulations enhance our understanding of applications such as assessing hydrogen storage efficiency and forecasting CO$_2$ sequestration processes in underground reservoirs.  Nevertheless, they are computationally expensive due to their mesoscopic nature. In addition, their stationary solutions are not guaranteed to be unique, so multiple runs with different initial conditions must be performed to ensure sufficient sample coverage. These factors complicate the task of obtaining  representative and reliable forecasts. To overcome the high computational cost hurdle, we propose a hybrid method that couples generative diffusion models and physics-based modeling. Upon training a generative model, we synthesize samples that serve as the initial conditions for physics-based simulations. We measure the relaxation time (to stationary solutions) of  the simulations, which serves as a validation metric and early-stopping criterion. Our numerical experiments revealed that the hybrid method exhibits a speed-up of up to 8.2 times compared to commonly used initialization methods. This finding offers compelling initial support that the proposed diffusion model-based hybrid scheme has potentials to significantly decrease the time required for convergence of numerical simulations without compromising the physical robustness.


\end{abstract}

\section{Motivation}
Hydrogen storage and CO$_2$ sequestration within the subsurface stand out as promising solutions for large-scale emission reductions, contributing significantly to climate change mitigation efforts \citep{heinemann2021enabling, zivar2021underground, bachu2008co2}. Thus, there is a growing interest in understanding the fluid dynamics of these processes better. When fluids (such as H$_2$ or CO$_2$) are injected into the subsurface, they undergo complex interactions with pre-existing fluids and the host rock. In addition, the presence of fractures—acting as preferential pathways for flow—further complicates this scenario, posing potential risks to the effectiveness of underground storage systems \citep{fitts2013caprock, guiltinan2021two, ting2022using}. 

Pore-scale simulations of multiphase flow provide an accurate picture of how fluids travel through underground reservoirs and capture their stable configurations given flow paths \citep{blunt2013pore, blunt2017multiphase}. However, scaling these simulations to the field level is impractical, primarily as simulating larger and more intricate samples entails significantly larger computational costs. In addition, the existence of multiple solutions for identical geometries and the requirement for numerous simulations to deduce constitutive relationships pose substantial challenges.

Diffusion models are powerful generative models capable of approximating high-dimensional data distributions \citep{ho2020denoising, sohl2015deep}. We hypothesize that  diffusion models can learn and synthesize near-optimal fluid configurations for pre-specified geometries, addressing the both physical complexities and computational issues inherent in the pore-scale modeling. In this study, we demonstrate the capabilities of diffusion models for generating multiphase fluid configurations. By introducing geometric conditioning to the model and assessing its performance with both computer vision metrics and numerical-solver-based metrics, the findings affirm the physical robustness and computational efficiency of our approach with diffusion models.

\section{Data: Fractures and Multiphase Simulations}
To train our model, we generate a diverse dataset of 1,000 2D synthetic fractures (shown in Figure \ref{fig:t-SNE}) using \texttt{pySimFrac}, a Python-based fracture surface generator \cite{guiltinan2023pysimfrac}. We ran two-phase steady-state fluid simulations using the MP-LBM library \citep{santos2022mplbm}. For each simulation, the two fluids are initialized at different ratios (from 20\% to 60\%) and are driven forward by an external force. The simulations iterate until they converge, indicated by a minimum in energy. The converged simulation depicts the most conductive pathways for each sample given the solid geometry and the initial ratios of fluid. It is worth noting that these paths may not be unique. The average simulation takes around 1M iterations in time (time-steps) to converge. Examples of converged simulations are shown in Figure \ref{fig:DNS_results}.


\section{Denoising Diffusion Probabilistic Models with Geometric Conditioning}
\begin{figure}[!t]
    \centering
    \includegraphics[scale=0.4]{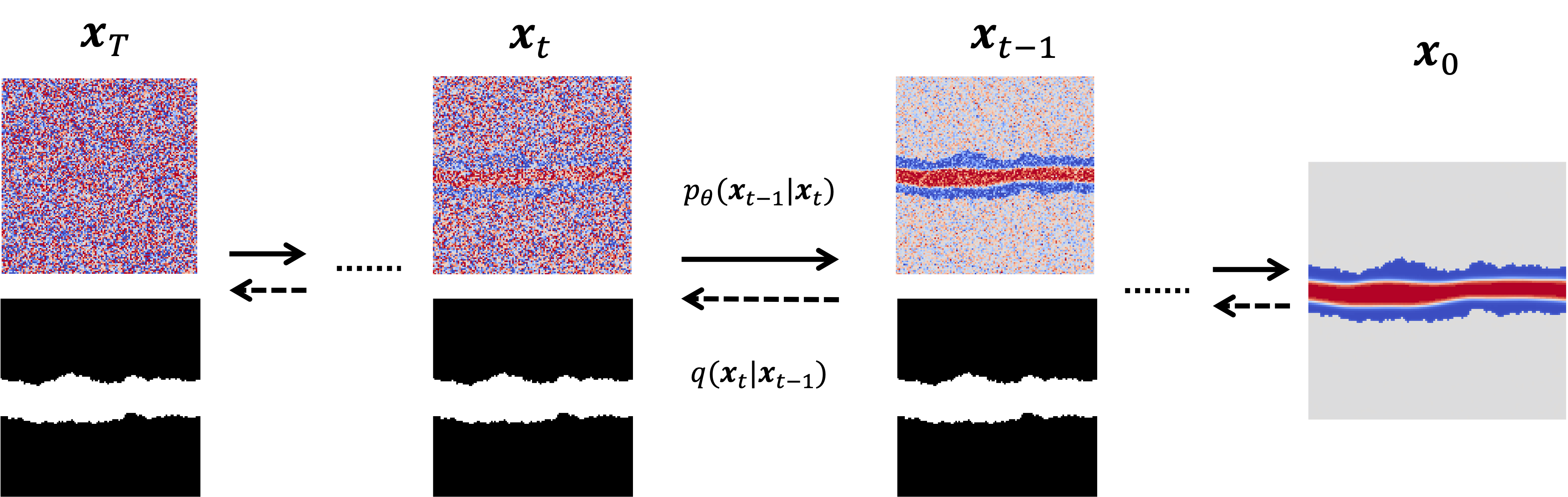}
    \caption{Diagram depicting the diffusion process and the integration of a geometry channel to guide fluid configurations within predefined geometric constraints.}
    \label{fig:diffusion_process}
\end{figure}

We employ Denoising Diffusion Probabilistic Models (DDPMs) \citep{ho2020denoising} to synthesize stable fluid configurations in fractured media. A forward Markov chain is defined to perturb samples drawn from the data distribution, \(\mathbf{x_0} \sim q(\mathbf{x_0})\), towards a limiting isotropic Gaussian distribution. The conditional mean of the reverse-time process were derived \cite{sohl2015deep,ho2020denoising} as the training target of a U-Net, which is used to propagate samples drawn from the limiting distribution back to the data distribution after training, generating novel samples. 
Variational inference (VI) is used to train the diffusion model via minimizing the evidence lower bound (ELBO), a lower bound of the Bayes evidence of the model \citep{sohl2015deep,ho2020denoising}. \citet{ho2020denoising} showed that VI can be effectively performed by training the model to predict the noise, $\bm{\epsilon}$, that is added to $\bm{x}_t$, given a noisy image $\bm{x}_t$, via the loss function 
\begin{equation}
    \mathcal{L}(\theta) = \mathbb{E}_{t \sim [1,T], \bm{x}_0, \bm{\epsilon}} [\| \bm{\epsilon} - \bm{\epsilon}_\theta (\bm{x}_t, t) \|^{2}]
\end{equation}
In the realm of fluid configurations, the geometry of flow paths significantly influences the configurations of fluids. Thus, for practical applications, the model needs to demonstrate its ability to generate fluid configurations that align with desired geometries rather than adopting arbitrary shapes. This task is similar to inpainting in computer vision \cite{yeh2017semantic, yu2018generative}, which requires the model to synthesize images in unknown areas that are both realistic and consistent with the surrounding background. One unique challenge in our inpainting tasks lies in the fracture geometries that feature sharp and irregular shapes, unlike the rectangular or smoothly varying unknown regions often seen in previous studies \citep{latt2021palabos, lugmayr2022repaint}.

To ensure that our model generates fluid configurations that adhere to given geometries (i.e., not solely based on random Gaussian noise), we introduce an auxiliary binary image $\bm{G} \in \{0,1\}^{H \times W}$ to encode the fracture geometry. Here, $H$ is the height and $W$ is the width of the image:
\begin{equation}
\bm{G}(i,j) = 
\begin{cases} 
1, & \text{if pixel at } (i,j) \text{ is part of the solid} \\
0, & \text{otherwise}
\end{cases}
\end{equation}
Figure \ref{fig:diffusion_process} shows a schematic diagram of the diffusion process of our model, which generates denoised images ($\bm{x}_{t-1}$) given $\bm{x}_t$ \emph{and} $\bm{G}$. By providing this additional channel in every denoising step during the reverse process, we give additional information for the model to synthesize images given a geometry. After training the model, we can feed an out-of-sample fracture geometry input ($\bm{G}$) and a sample of the isotropic Gaussian random variable ($\bm{x}_T$) to generate fluid configurations in the desired domain similar to input geometry for numerical simulation procedures detailed below.

\section{Results}
\subsection{Training and Model Performance}
We employ two evaluation metrics to assess the performance of our model during training. The first one is the Mean Squared Error (MSE), a metric in computer vision to quantify the discrepancy between the predicted and the ground-truth images. Although the MSE loss in the U-Net provides its denoising capability, evaluating the optimal model for the physical domain is challenging. Thus, we apply a second metric that focuses on the number of iterations required for Direct Numerical Simulation (DNS) to converge, starting from the initial configuration given by the diffusion model. The underlying rationale for this metric is grounded in the functionality of numerical simulators that approximate solutions of partial differential equations (PDE) through iterative methods. Consequently, if the fluid configuration generated by our diffusion model closely approximates the numerical PDE solution, the number of iterations needed for DNS convergence should be reduced.

\begin{figure}
    \centering
    \begin{subfigure}{0.3\textwidth}
        \includegraphics[width=\textwidth]{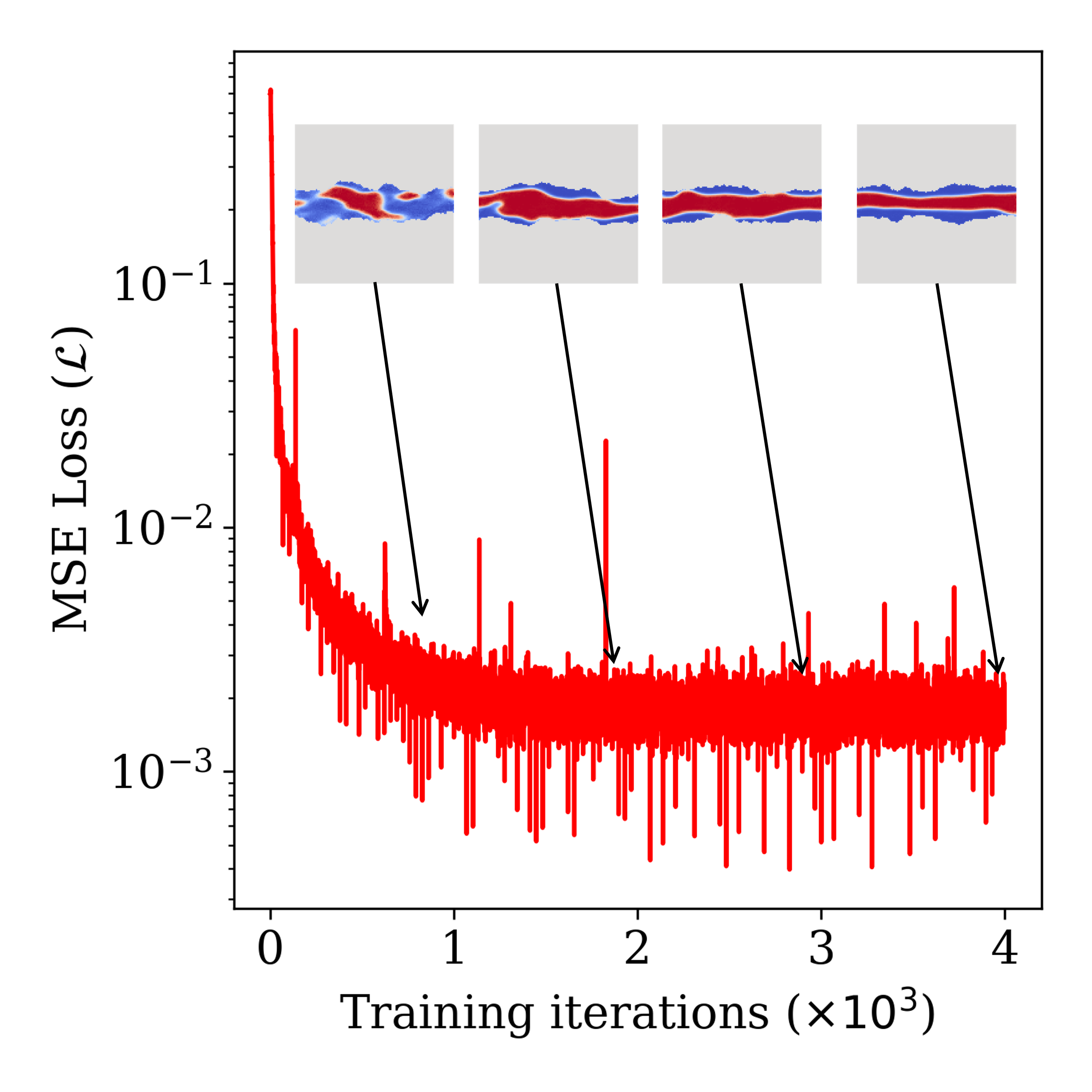}
        \caption{Evolution of MSE loss across iterations, showcasing samples at 1K, 2K, 3K, and 4K training iterations.}
        \label{fig:Loss_Change}
    \end{subfigure}
    \hfill
    \begin{subfigure}{0.3\textwidth}
        \includegraphics[width=\textwidth]{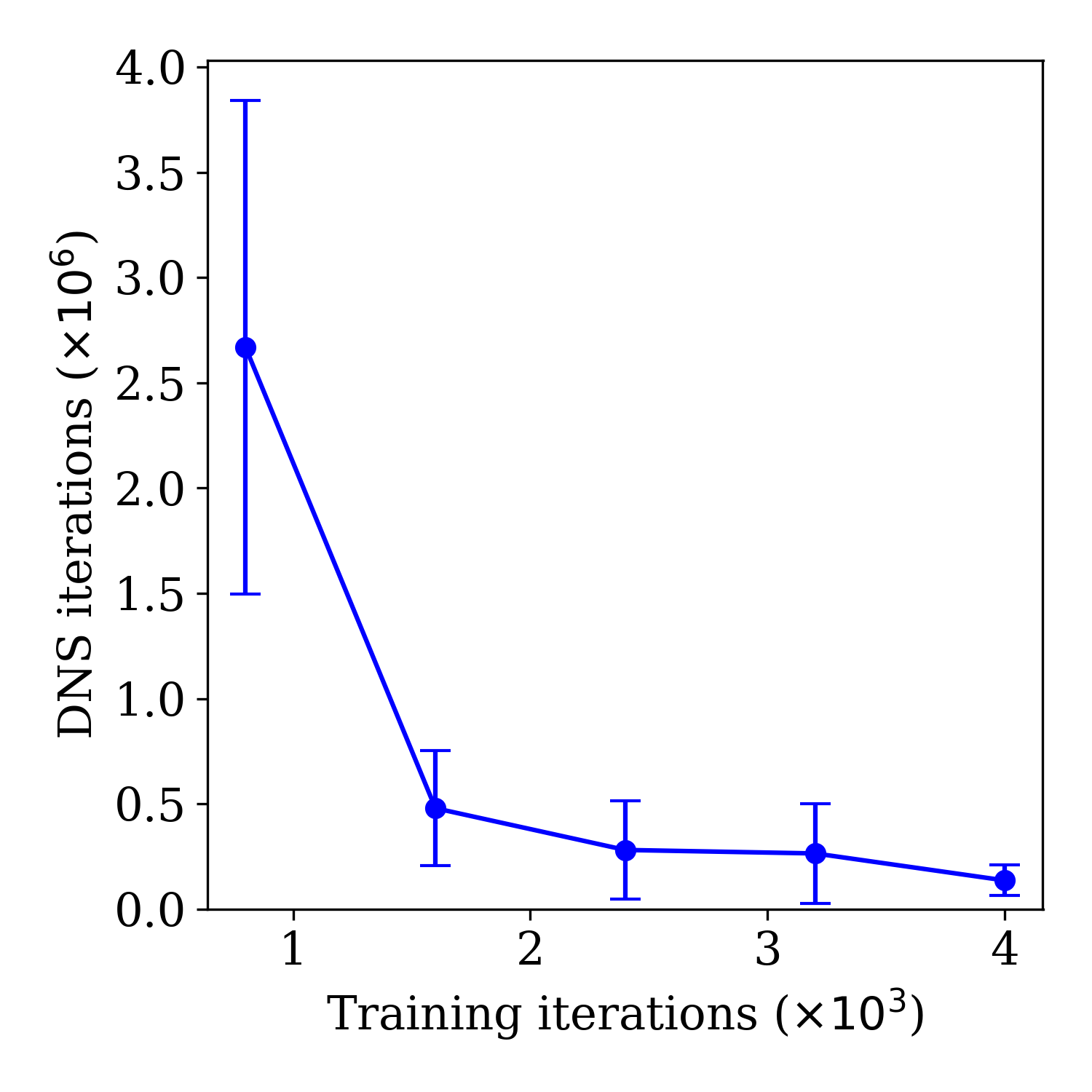}
        \caption{Variation in iterations needed for DNS convergence, with the mean and standard deviation ($\mu \pm 0.5 \sigma$)}
        \label{fig:Iter_Change}
    \end{subfigure}
    \hfill
    \begin{subfigure}{0.3\textwidth}
        \includegraphics[width=\textwidth]{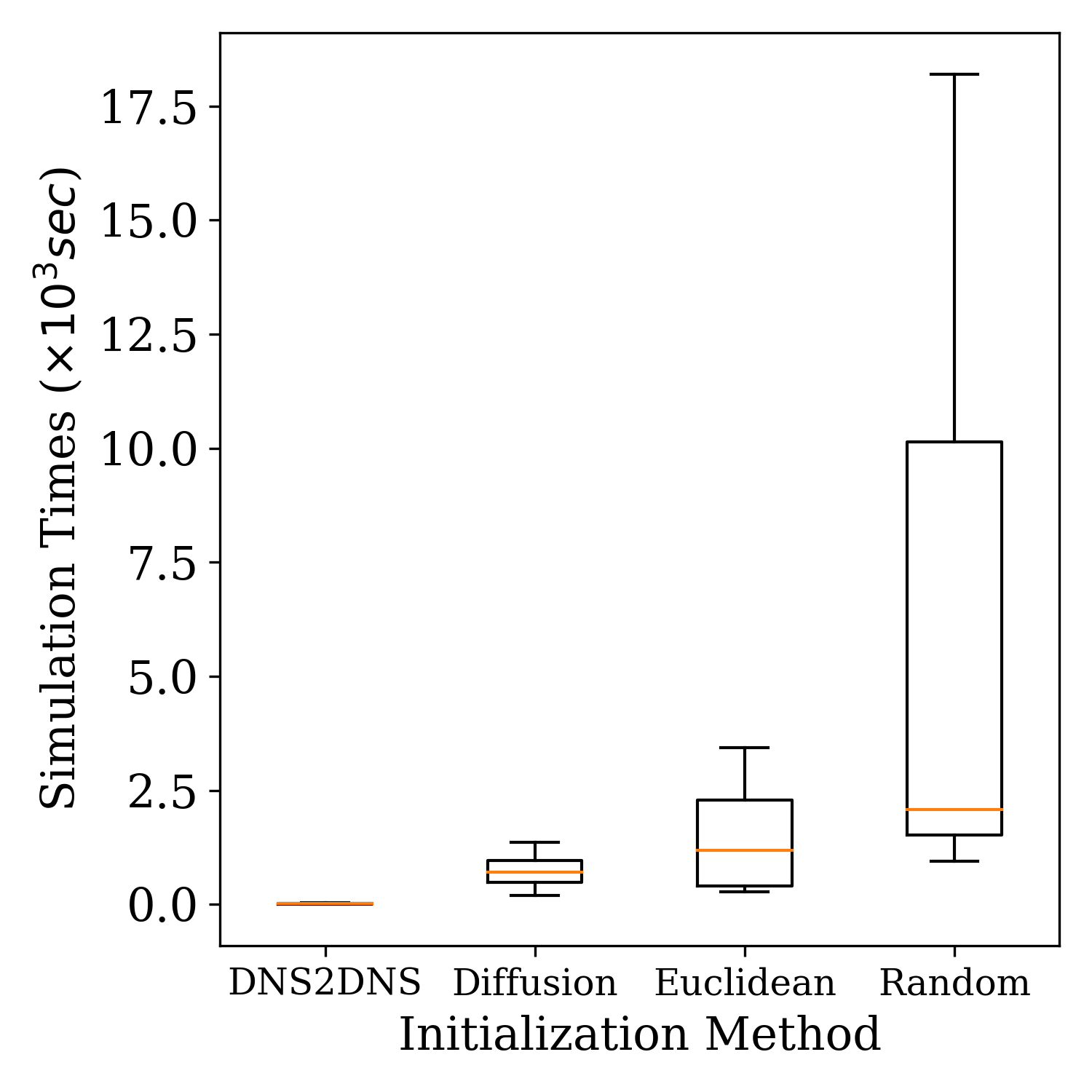}
        \caption{Simulation times with different initialization methods: DNS, Diffusion, Euclidean distance, and Random initialization.}
        \label{fig:Simulation_t}
    \end{subfigure}    
    \caption{Summary of results: evolution of MSE loss, variation in iteration number for DNS convergence, and comparison of simulation times across different initialization methods.}
    \label{fig:Results}
\end{figure}

Throughout training, we generate fluid configurations for unseen fracture geometries at 800-iteration intervals. These configurations serve as initial conditions for multiphase flow simulations, and the requisite iterations for DNS convergence are recorded. The convergence is monitored through relative kinetic energy \cite{latt2021palabos}, represented as $({1}/{2}) \left\lVert {\partial \vec{v}}/{\partial t} \right\rVert^2$, where ${\partial \vec{v}}/{\partial t}$ signifies the time rate of change of the velocity vector. The simulation is deemed converged once the integration of this quantity over the simulation domain is below a threshold $\kappa$ (set to $5 \times 10^{-5}$ in our study). Figure \ref{fig:Loss_Change} and Figure \ref{fig:Iter_Change} illustrate the evolution of the MSE loss and the DNS convergence iterations over training time. Notably, the MSE loss starts from a high value and plateaus around 2K training iterations, indicating effective denoising for the training dataset. Moreover, we observe a significant reduction in the number of iterations required for DNS convergence. Specifically, the average number of iterations dropped from $2.588 \times 10^6$ to $0.137\times 10^6$, a reduction of approximately 95\%. The standard deviation also decreased by 95\%, further emphasizing the model's increasing stability. These results suggest that our diffusion model is not only able to capture the visual similarity of the training dataset but also learn to generate physical configurations conditioned on the fracture geometry.

\subsection{Computational Cost Comparison}
Upon completion of the training, we explore the computational advantages of utilizing the hybrid approach---deploying the diffusion model for fluid initialization---by analyzing the requisite simulation time for convergence. To validate our strategy, we benchmark against four fluid initialization methods: (1) DNS solution, (2) diffusion-based initialization, (3) Euclidean distance-based initialization, and (4) random initialization. Detailed descriptions and examples of each method are provided in the Appendix \ref{section:Appen_C}. Briefly, the DNS solution-based initialization method and the random-based initialization method represent the lower and upper bounds of convergence time, respectively, while the Euclidean-based initialization is a reasonable approach, grounded in the domain knowledge that invading fluid tends to occupy pores distant from solid walls \cite{blunt2017multiphase}.



\begin{table}
  \caption{Computational times for different initialization methods, based on a 50-core numerical simulator. The total time for our diffusion-based simulation is $858.8$s (image generation time ($120$s) + numerical simulation time ($738.8$s))}
  \label{tab:comp_times}
  \centering
  \begin{tabular}{ccc}
    \toprule
    Initialization Method & Mean Time (s) & Standard Deviation (s) \\
    \midrule
    DNS Solution & 14.2 & 7.3  \\
    Diffusion & 738.8 & 425.5  \\
    Euclidean & 1515.3 & 1281.2 \\
    Random & 7075.0 & 7880.2 \\
    \bottomrule
  \end{tabular}
\end{table}

Figure \ref{fig:Simulation_t} and Table \ref{tab:comp_times} represent the simulation times for convergence for each initialization method. In particular, employing the diffusion output as initialization necessitates a mean simulation time of approximately 738.8 seconds for convergence. Considering the additional diffusion generation time of 120 seconds, the diffusion-based approach's aggregate computational time is approximately 858.8 seconds. Given these results, the diffusion-based approach exhibits a computational speed-up factor ranging from $1.76$ times to $8.24$ times compared to Euclidean and random initialization methods, respectively. This quantitative interpretation underscores the superior computational efficiency of the diffusion-based approach among the assessed initialization methods. The pronounced reduction in simulation times, particularly in comparison to the random initialization method, highlights the potential and practicality of integrating diffusion models for fluid initialization in computational fluid dynamics simulations.


\section{Conclusion}
We utilized a diffusion model to synthesize fluid configurations tailored to specific fracture geometries. Our method effectively addresses the challenges posed by multiple solutions due to varying saturation levels, as well as the high computational costs commonly associated with the pore-scale simulations. To evaluate the model performance, we introduced a domain-specific metric: the number of iterations required for convergence in DNS. Our results reveal that this metric significantly improves as the training progresses, underscoring the model's ability to learn the underlying physics of fluid configurations, not just their visual attributes. In particular, we validated our model's performance using an unseen dataset, reinforcing its robustness and generalizability. Moreover, we quantified the computational advantages of our hybrid approach that combines diffusion model-based initialization with DNS. Our findings indicate that this method can accelerate simulations by factors ranging from 1.76 to 8.24 times compared to traditional methods, offering substantial computational cost savings. This finding offers compelling initial support that the proposed diffusion model-based hybrid scheme has the potential to significantly decrease the computational cost of numerical simulations without compromising the physical robustness.

\section*{Acknowledgements}
J.C. thanks the Applied Machine Learning (AML) summer program and Center for Nonlinear Studies (CNLS) at Los Alamos National Laboratory (LANL) for the mentoring and the fellowship received. J.E.S acknowledges the support by the Laboratory Directed Research \& Development project ``Diffusion Modeling with Physical Constraints for Scientific Data (20240074ER)''.

\bibliographystyle{plainnat.bst}
\bibliography{reference.bib}

\newpage
\appendix
\counterwithin{figure}{section}
\counterwithin{table}{section}

\section{Synthetic fracture dataset}
We employed \texttt{pySimFrac} to generate a diverse set of synthetic fractures using the spectral method, adjusting the Hurst exponent ($H$), Anisotropy ratio ($r_a$), and Roughness ($\sigma$). Here, we briefly introduce the procedure for generating the fracture dataset. Readers interested in details can refer to the literature \cite{guiltinan2023pysimfrac}. The process starts by calculating the wavenumber increments $\Delta q_i$ and defining wavenumbers $q_x$ and $q_y$ for each grid point in Fourier space. The Fourier spectrum function $F(q_x, q_y) = A(q) [\cos(\phi) + i \sin(\phi)]$ is defined, where the amplitude $A(q) = q^{-1-H}$, and $q$ is computed by taking into account the anisotropy ratio. The phase spectra $\phi$ are generated randomly for the upper and lower surfaces of the fracture, respectively. Upon obtaining $F(q_x, q_y)$, we use the inverse Fourier transform to get $f(x, y)$ and adjust the roughness to obtain $f_{\text{rescaled}}(x, y)$. The complete fracture geometry is formed by the difference between the top and bottom surfaces of the fracture, formulated as $f(x, y) = f_{\text{top}}(x, y) - f_{\text{bottom}}(x, y)$. To introduce diversity, we define the range of three variables as (1) \(H \in [0.7, 0.9]\), (2) \(r_a \in [0.1, 0.9]\), and (3) \(\alpha \in [2, 6]\), resulting in 1000 diverse synthetic fractures.

\begin{figure}[h!]
  \centering
  \includegraphics[width=1\linewidth]{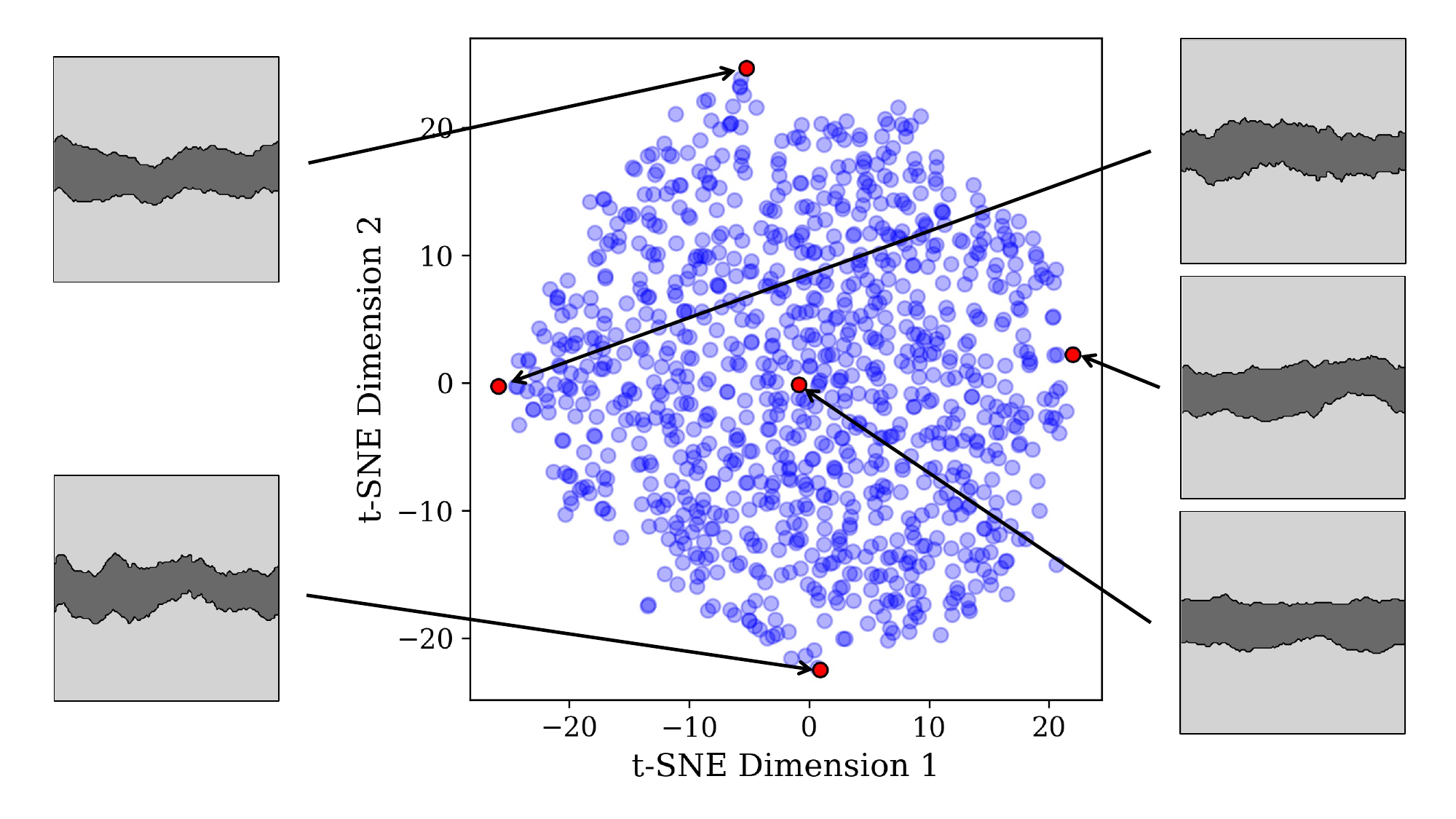}
  \caption{t-SNE visualization of the synthetic fracture dataset, showing its diversity with representative fracture examples}
  \label{fig:t-SNE}    
\end{figure}

To visualize the diversity within our dataset, we employed t-SNE (t-Distributed Stochastic Neighbor Embedding), a technique for reducing dimensionality and visualizing high-dimensional data structures in a lower-dimensional space \cite{van2008visualizing}, as shown in Figure \ref{fig:t-SNE}. This technique visualizes the representation of similar fracture geometries closely together and dissimilar ones further apart in a 2D plane, thereby providing insights into the variability and complexity within the dataset. The resulting t-SNE plot serves as a qualitative validation of the diversity of our fracture geometries, illustrating the broad spectrum of variations encompassed in the dataset.

\section{Examples of multiphase fluid configurations from numerical simulations}
\begin{figure}[H]
  \centering
  \includegraphics[width=1.0\linewidth]{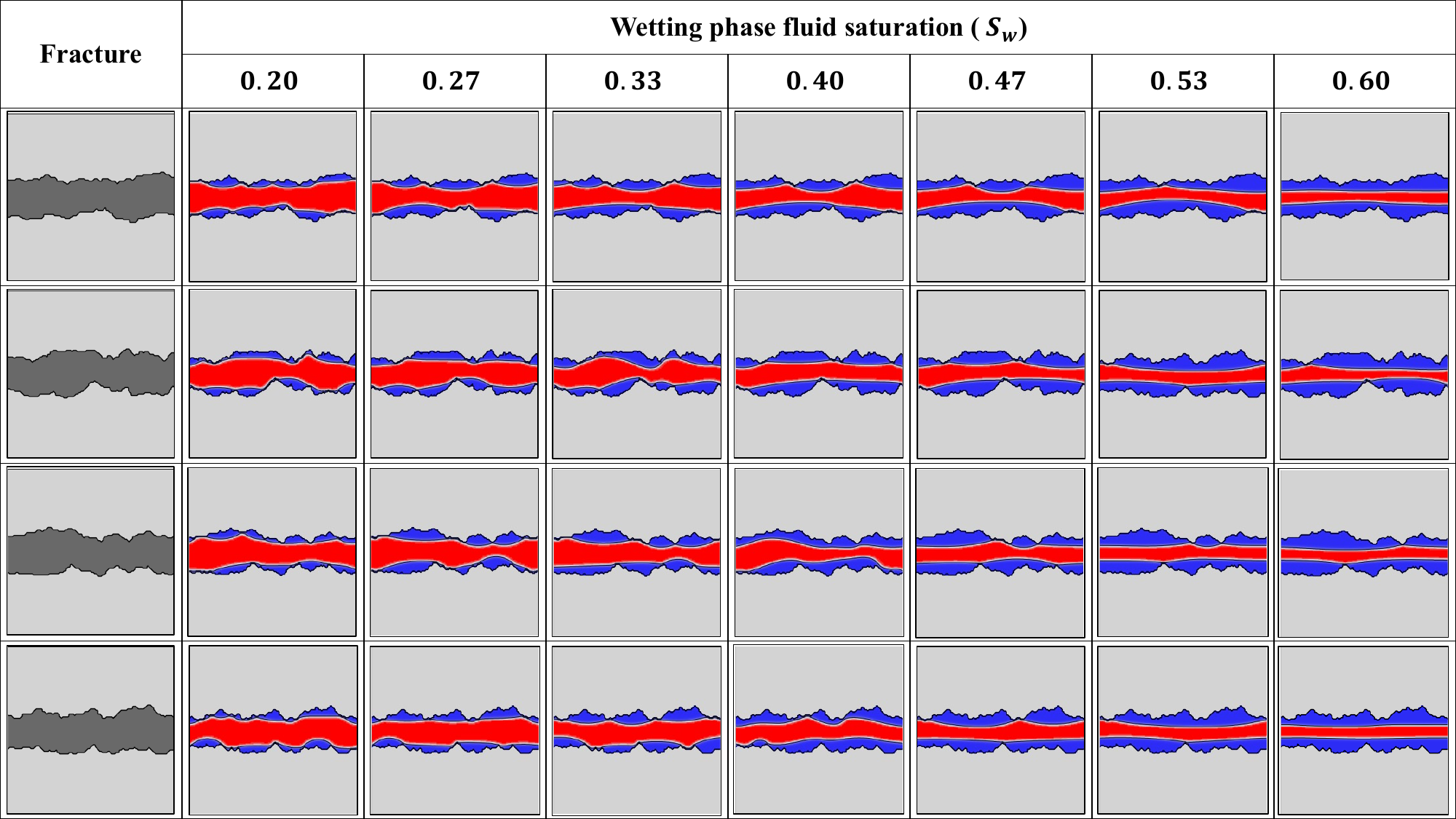}
  \caption{Fluid configurations obtained from simulation. Solid regions are represented in grey, and the two fluids in red and blue. The left and right boundaries are periodic. The fluids' behavior is complex, not solely determined by Euclidean distance or simple volumetric changes based on saturations. Instead, the fluids navigate through the complex geometries and interact with the solid surface in a manner that minimizes the system's free energy.}
  \label{fig:DNS_results}  
\end{figure}

\section{Two-phase fluid initialization methods}
In multiphase simulations utilizing the Shan-Chen model, initializing fluid phases with defined saturations and fluid paths is required. Various fluid initialization strategies are explored across distinct geometries to assess the efficacy of the diffusion-based initialization method. In particular, the density of the pre-existing fluid phase is defined to indicate wetting phase saturation and initial locations of the two-phase fluids. The density values span continuously from -0.4 to 2.1, where -0.4 corresponds to the wetting phase (pre-existing fluid), and 2.1 signifies the non-wetting phase (invading fluid).

\begin{figure}
    \centering
    \begin{subfigure}{0.24\textwidth}
        \includegraphics[width=\textwidth]{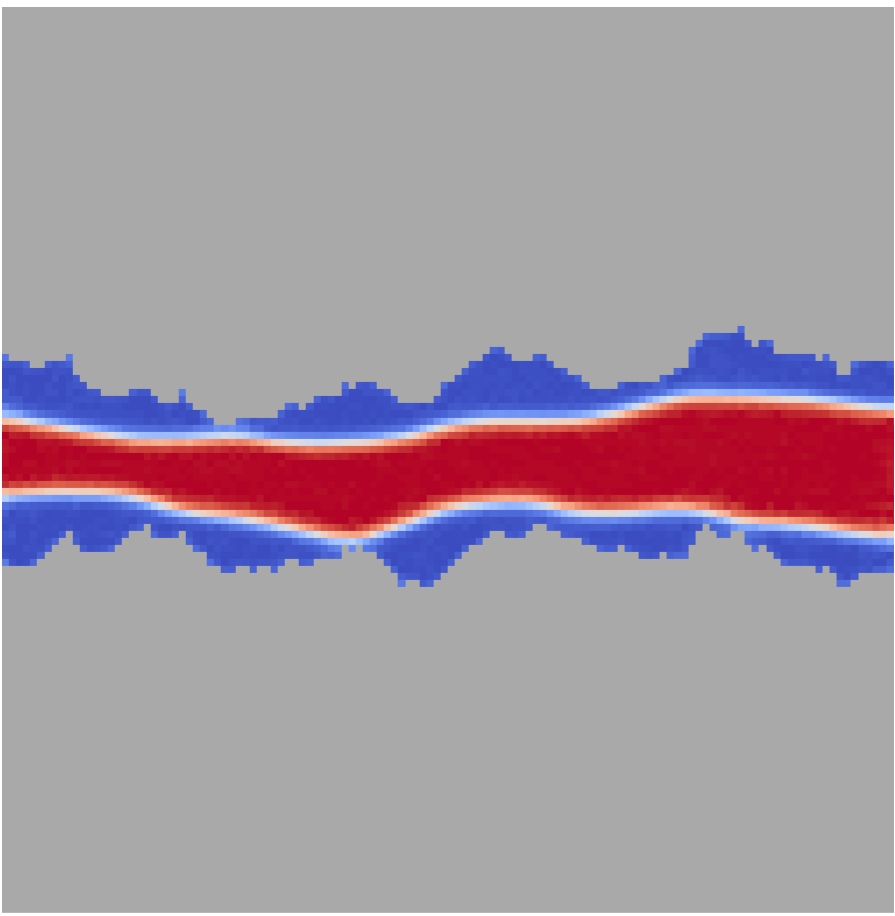}
        \caption{Diffusion}
        \label{fig:diffusion_init}
    \end{subfigure}
    \hfill
    \begin{subfigure}{0.24\textwidth}
        \includegraphics[width=\textwidth]{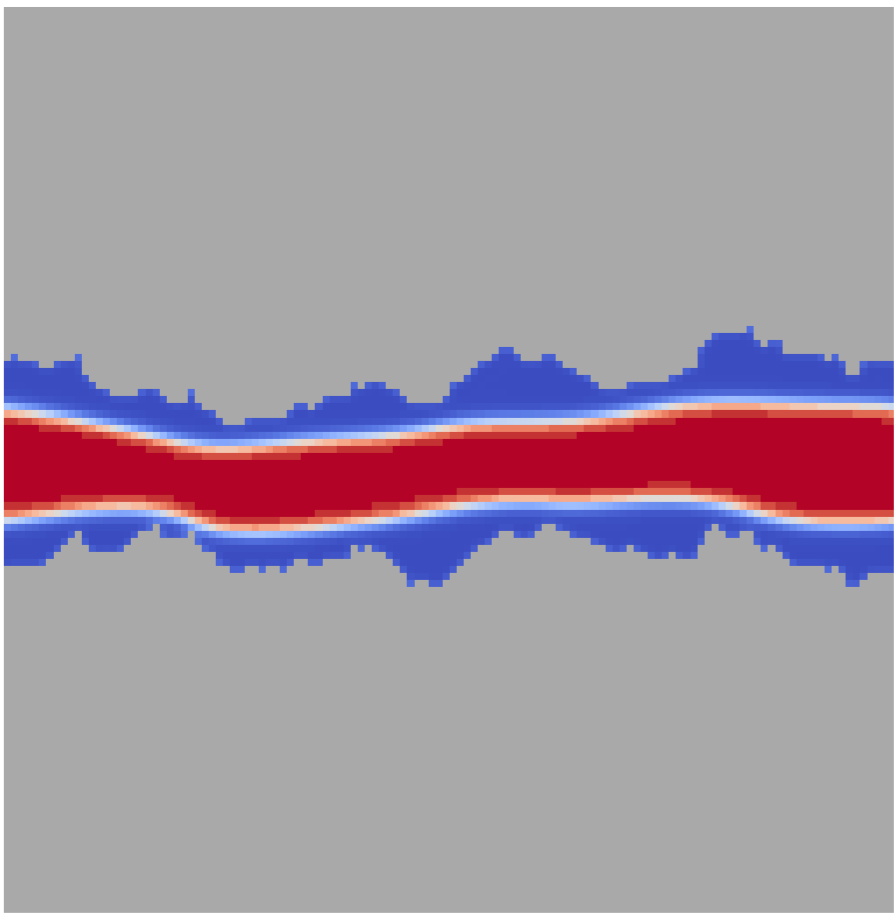}
        \caption{DNS solution}
        \label{fig:DNS_init}
    \end{subfigure}
    \hfill
    \begin{subfigure}{0.24\textwidth}
        \includegraphics[width=\textwidth]{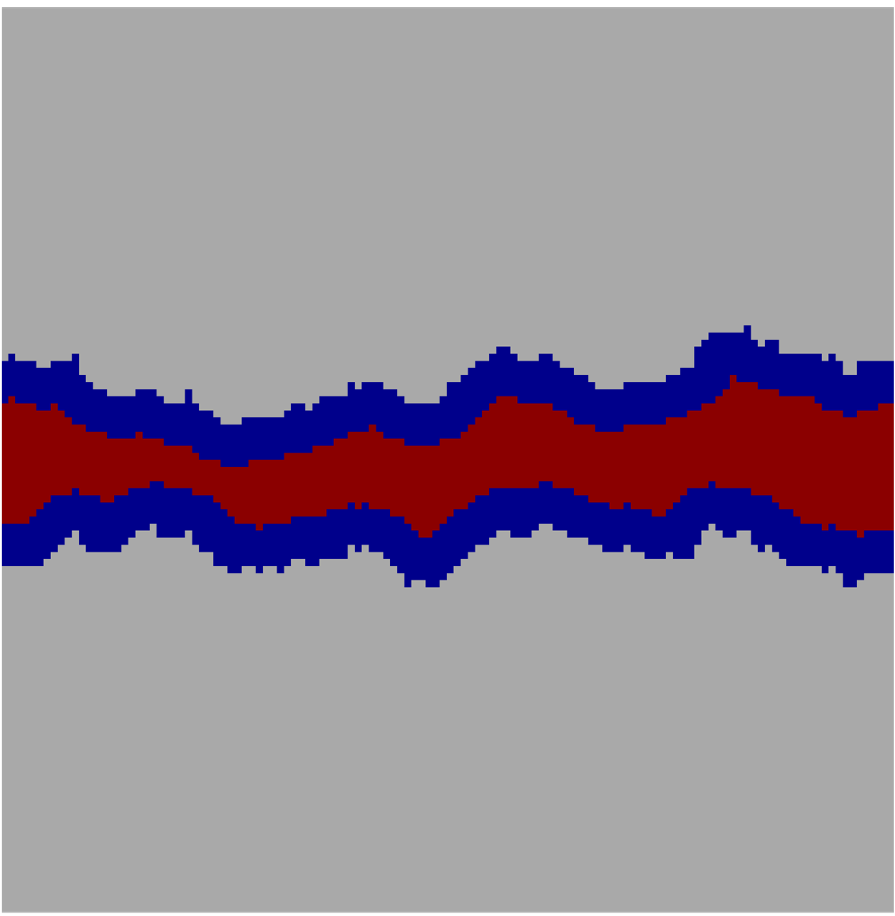}
        \caption{Euclidean}
        \label{fig:Euclidean_init}
    \end{subfigure}    
    \hfill
    \begin{subfigure}{0.245\textwidth}
        \includegraphics[width=\textwidth]{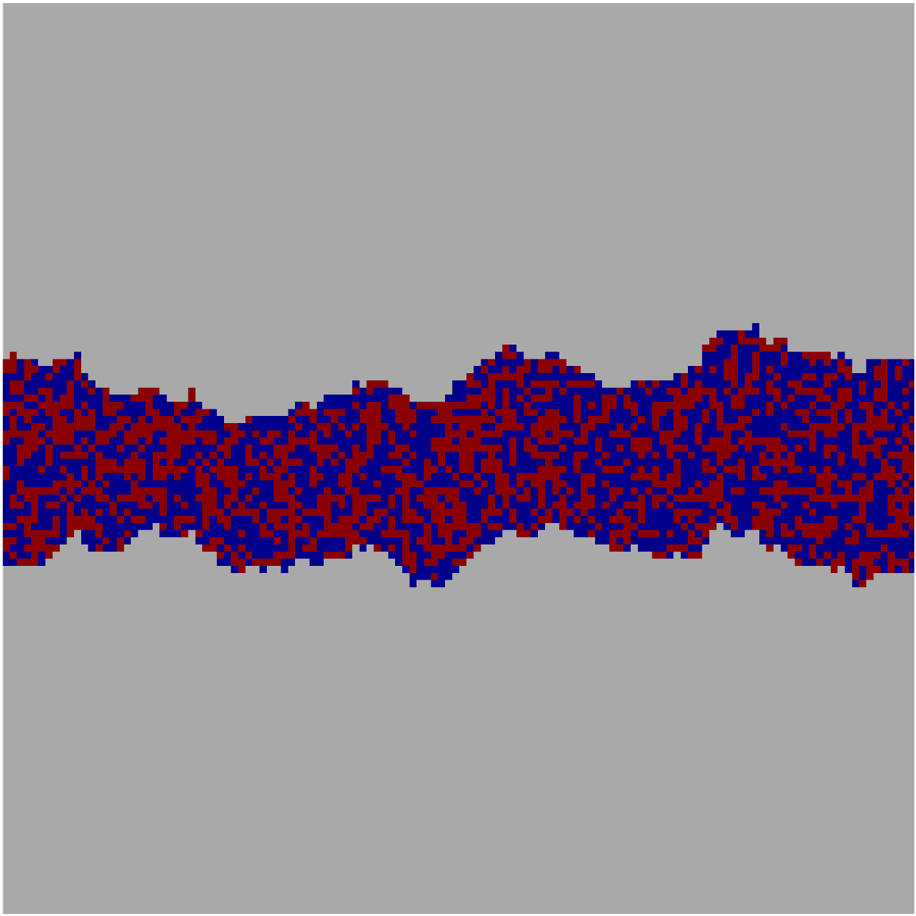}
        \caption{Random}
        \label{fig:Random_init}
    \end{subfigure}        
    \caption{Comparison of fluid configurations initialized by different methods within the simulation domain for multiphase simulations utilizing the Shan-Chen model. The Diffusion-Based Initialization synthesizes continuous values similar to the DNS solution. In contrast, the Euclidean Distance-Based and Random Initialization methods produce discrete phase representations.}
    \label{fig:init_comp}
\end{figure}

Figure \ref{fig:init_comp} presents examples of fluid configurations initialized by the different methods within the simulation domain. First, the \textit{Diffusion-Based Initialization} method leverages generated images to determine fluid configurations. Second, the \textit{Numerical Solution-Based Initialization} employs numerical solutions to initialize two-phase arrangements obtained from the simulation results of other initialization methods. Third, the \textit{Euclidean Distance-Based Initialization} identifies the non-wetting region in the flow path using provided saturations and positions the non-wetting fluid away from the solid boundary, based on the understanding that invading fluid tends to inhabit pores distant from solid walls \cite{blunt2017multiphase}. Lastly, the \textit{Random Initialization} method arbitrarily selects the positions of two-phase fluids in accordance with their specified saturations and flow paths. A notable advantage of the Diffusion-Based Initialization is that is able to initiate continuous density values, which is essential for considering local capillary pressure differences like the physics-based solution fields. In contrast, the Euclidean Distance-Based and Random Initialization methods can only produce discrete phases.

\label{section:Appen_C}

\end{document}